\documentclass[a4paper,10pt, compsoc, onecolumn]{IEEEtran}

\usepackage[utf8]{inputenc}
\usepackage{amsmath, amssymb}
\usepackage{array}
\usepackage[T1]{fontenc}
\usepackage{amssymb,amsfonts}
\usepackage{mathrsfs}
\usepackage{enumerate}
\usepackage{color}
\usepackage{geometry}
\newcommand{\RR}{\mathbf{R}}
\newcommand{\NN}{\mathbf{N}}
\usepackage{amsthm}
\usepackage{algorithm}

\usepackage{algpseudocode}

\usepackage{titleps}
\usepackage[numbers]{natbib}
\newcommand\add[1]{#1}

\renewcommand{\emph}{\textit}

\usepackage{graphicx}

\newpagestyle{main}{
  \sethead{}{}{}
  \setfoot{}
    {\thepage}
    {}
  \setheadrule{0pt}
  \setfootrule{0pt}
}

\begin{document}
\title{Virtual Function Placement for Service Chaining with Partial Orders and Anti-Affinity Rules}

\author{  \IEEEauthorblockN{Zaid Allybokus\IEEEauthorrefmark{1}, Nancy Perrot\IEEEauthorrefmark{2}, Jérémie Leguay\IEEEauthorrefmark{1}, Lorenzo Maggi\IEEEauthorrefmark{1}, Eric Gourdin\IEEEauthorrefmark{2}}
    \IEEEauthorblockA{\IEEEauthorrefmark{1}Huawei Technologies, France Research Center
    \\\{zaid.allybokus, jeremie.leguay, lorenzo.maggi\}@huawei.com}
    \IEEEauthorblockA{\IEEEauthorrefmark{2}Orange Labs, France
    \\\{nancy.perrot, eric.gourdin\}@orange.com}
}

\maketitle
\IEEEdisplaynontitleabstractindextext
\thispagestyle{main}
\pagestyle{main}
\begin{abstract}
Software Defined Networking and Network Function Virtualization are two paradigms that offer flexible software-based network management. Service providers are instantiating Virtualized Network Functions - e.g., firewalls, DPIs, gateways - to highly facilitate the deployment and reconfiguration of network services with reduced time-to-value. They employ Service Function Chaining technologies to dynamically reconfigure network paths traversing physical and virtual network functions. Providing a cost-efficient virtual function deployment over the network for a set of service chains is a key technical challenge for service providers, and this problem has recently caught much attention from both Industry and Academia. In this paper, we propose a formulation of this problem as an Integer Linear Program that allows one to find the best feasible paths and virtual function placement for a set of services with respect to a total financial cost, while taking into account the (total or partial) order constraints for Service Function Chains of each service and other constraints such as end-to-end latency, anti-affinity rules between network functions on the same physical node and resource limitations in terms of network and processing capacities. Furthermore, we propose a heuristic algorithm based on a linear relaxation of the problem that performs close to optimum for large scale instances. \\

\noindent\textbf{Keywords:} Software-Defined Networking; Network Function Virtualization; Service Function Chaining; Anti-Affinity rules; Partial order; Integer Linear Programming.

\end{abstract}


\section{Introduction}

In addition to packet forwarding equipment such as switches and routers, service provider networks commonly include middleboxes, also called \emph{Service Functions} (SFs), such as firewalls, DPIs, web proxies, media gateways, etc. These SFs are traditionally provided with a specific hardware. Their deployment and reconfiguration thus suggest manual interventions in order to connect them to the network via wired cables, which is costly and time-consuming. Nowadays, SFs are delivered as Virtual Machines (VM) that can run on off-the-shelf servers thanks to Network Function Virtualization (NFV) technologies~\cite{li2015software}.  
They have yet to be improved to provide comparable performances: they still usually represent less effective versions of their hardware peers in terms of  packet processing capacity and speed. However, their reduced cost and generic hardware compatibility allow them to be duplicated easily and to adapt quickly to evolving traffic. In addition to this transformation, Software-Defined Networking (SDN) offers the unprecedented possibility to separate the control plane from the data plane, providing high programmability in the management and control of individual flows. Consequently, SDN, along with NFV, are two elegant paradigms that offer flexible software-based network management.

One major challenge for service providers is to deploy \emph{Service Function Chains} (SFC), where a customer demand has to traverse SFs in a pre-defined, flow-specific order. 

For instance, a link connecting two enterprise sites may need to pass through a firewall, a DPI, and a web proxy. 
While SFCs have been initially defined as a set of network functions that must be traversed in a given order, this model can be refined to better meet service providers needs and optimize resources. 
While order constraints may impose a given function, say a classifier, to be traversed prior to a WAN optimizer, some functions (e.g., a firewall) may not have ordering constraints. Recent studies~\cite{Addis15} have showed that this flexibility optimizes resource usage as it may reduce the number of activated functions and decrease the network footprint. 

For this reason, we believe the possibility to set partial order constraints is an interesting feature. In addition, as mentioned in requirements~\cite{etsi_doc} from the ETSI NFV standardization group, the so called \emph{anti-affinity} rules are generally used to reduce the impact of an infrastructure failure. More specifically, anti-affinity rules prevent certain Virtual Network Functions (VNF) within the same SFC from sharing the same physical resources. This increases the robustness of the service against one node failure at the same time.

Motivated by this, in this paper we address the joint VNF Placement and Routing (VNF-PR) problem for Service Function Chaining in the case of partially ordered SFCs satisfying the anti-affinity rules. To the best of our knowledge, the partial orders and anti-affinity rules we consider in this work have not been investigated before.

We first provide an Integer Linear Program (ILP) formulation that jointly searches for an optimal placement and an individual path for each service chain, while minimizing the overall cost of jointly traffic engineering and VNF instantiation. Our formulation is naturally flexible and permits one to relax orders easily. The unconstrained VNF-PR problem is known to be NP-hard~\cite{Cohen15} as it combines integer versions of the facility location and multi-commodity flow problems. Therefore, heuristics that well approximate the optimum in a reasonable amount of time are a must for service providers and carriers. We thus design a heuristic based on the linear relaxation of our ILP formulation and we show numerically its effectiveness. We also evaluate its performance on the realistic GEANT network topology, using the ILP as performance benchmark and by comparing it to a greedy algorithm that treats commodities in sequence.

The paper is organized as follows. We first present related works in  Section~\ref{relatedwork}. Section~\ref{problemstatement} states formally the problem while in Section~\ref{ilpformulation} we present the ILP formulation, providing the optimal solution to our problem. As a warm up, Section~\ref{greedyalgorithm} introduces the naive, greedy algorithm to tackle the problem. In turn, we propose in  Section \ref{heuristicdesign} a heuristic approach. Our numerical evaluation is then commented upon in Section~\ref{evaluation}, and finally, Section~\ref{conclusion} concludes the paper.
\vspace{5mm}

\section{Related Work}\label{relatedwork}

NFV leverages IT virtualization technology to consolidate many network services (network address translation, ciphering, firewalling, intrusion detection, caching, etc.) onto standard high volume servers which could be located at Points of Presence (POP), data centers or end user premises~\cite{mijumbi2015network}. 


The VNF-PR problem is to allocate server and network resources such that a given set of service chains can be deployed onto the network at minimum cost. A few papers~\cite{Addis15,mehraghdam2014specifying,moens2014vnf} have recently proposed ILP formulations considering constraints on end-to-end latency, network capacity and computing resources (e.g., maximum number of chains and functions, maximum amount of traffic). Moens et al.~\cite{moens2014vnf} consider hybrid chains that can include both physical and virtual resources. Mehraghdam et al.~\cite{mehraghdam2014specifying} define a grammar to characterize network chains in which one of the primitives is \textit{optional order} to relax order constraints between functions.

Nevertheless, this grammar is not part of their ILP formulation. The authors of~\cite{mehraghdam2014specifying} propose an exhaustive search to find all possible permutations and then, they solve the unconstrained ILP for each of them.
Addis et al.~\cite{Addis15} consider a formulation without any order constraints and that leverages traffic compression properties from multimedia gateways (e.g., video adaptation) or firewalls to reduce the network footprint. They propose a linearization of the model. 
Our main contribution to the problem formulation is to consider partial orders and anti-affinity rules.


A few algorithms have already been proposed to solve the VNF-PR problem with latency and resource constraints. Ghaznavi et al.~\cite{GhaznaviSAB16} propose a
local search heuristic to trade off between accuracy and speed. The minimum improvement at each step is used as a tunable parameter. Luizelli et al.~\cite{Luizelli15} use a binary search to find the fewest possible number of functions and solve a reduced ILP. Both papers consider service chains with a fully determined order of functions. Cohen et al.~\cite{Cohen15} study a bi-criteria approximation algorithm with a factor of 8 for the cost minimization objective function and a factor of 16 for the size constraints at the nodes. 
Beyond offline problems, deterministic online algorithms have been proposed for 
Service Function Chaining by Lukovszki et al.~\cite{lukovszki2014online} or for the incremental deployment of a single type of function~\cite{Lukovszki:2016} by the same authors.


Several research works~\cite{Fischer13} have proposed solutions to the \emph{Virtual Network Embedding} (VNE) problem, where several virtual networks have to be mapped onto a physical substrate. However, the problems of embedding virtual networks differ from the VNF-PR problem as virtual nodes of the same type \emph{but from the different requests} must share the same substrate node if it is cost efficient.
Some authors relate the VNF-PR problem with full order constraints to the VNE literature~\cite{chowdhury2009virtual, GhaznaviSAB16}, where SFCs are modeled as Virtual Network requests (VNr). Fuerst et al.~\cite{fuerst2013virtual} propose a pre-clustering algorithm that enables collocation of virtual nodes (or SFs) of \emph{the same} VNr (or SFC) onto the same substrate node. However, we argue that the VNE literature neither responds to the mutualization of virtual nodes (SFs) of the same type between different VNrs (SFCs) nor to the partial orders within the same SFC. As such, to the best of our knowledge, these two problems are separate.

Our contribution is twofold. We first propose an ILP formulation of the VNF-PR problem with partial order and anti-affinity constraints which we reinforce by adding redundant cuts. Second, we propose and evaluate a heuristic approach based on a linear relaxation to solve the problem.

\vspace{5mm}
\section{Problem Statement}\label{problemstatement}
We consider a network whose nodes, in addition to their traditional role of handling packet forwarding, are equipped with processors that allow them to support the VNFs that are required by the different connection requests, which we will now refer to as \emph{commodities}. They contain a limited number of cores that can each support a function instantiation. Each instantiation of those functions can handle multiple commodities up to a certain maximum rate (function capacity). If needed, functions can be replicated on the same node. However, this replication demands the dedication of an additional core. Each commodity is provided with an SFC that needs to be deployed along its chosen path. This list can be unordered, totally ordered, or partially ordered. A Service Function Path (SFP) is the physical path through an SFC instantiation that is chosen for a commodity. 

Instantiating a service function has a cost (which may contain the price for set up, license, and maintenance), which motivates the need to assign as many commodities as possible to active instances if they have enough capacity.

Moreover, routing traffic through physical links has a per-rate cost and motivates the need to preferably allocate shortest paths. 

The anti-affinity rules seek to enforce the reliability of a VNF structure. They forbid the assignment of certain VNFs to one commodity on the same location to ensure that SFPs do not entirely break down by a single server failure. 

\begin{figure*}[tb]
\centering
\includegraphics[width=\textwidth]{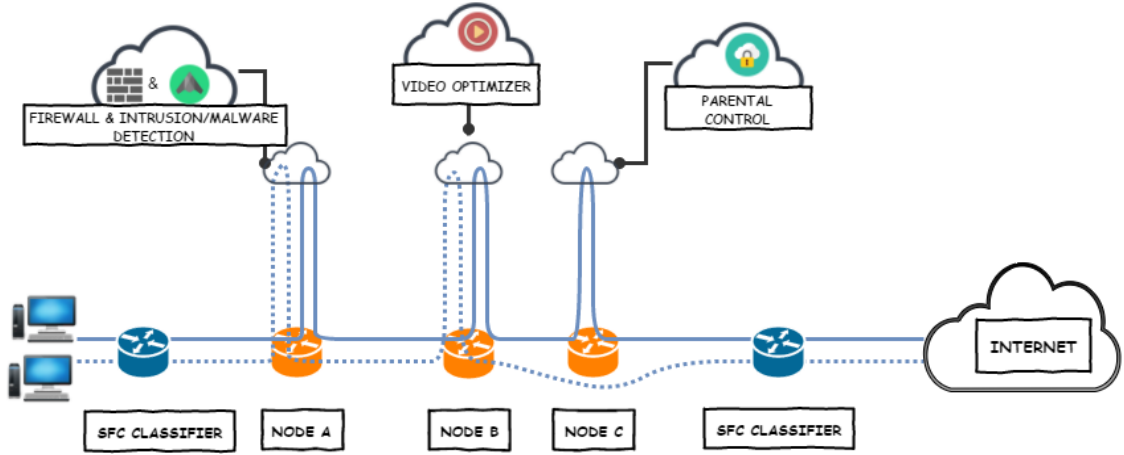}
\caption{Service Function Chaining use case. In this example, two SFCs share VNFs deployed over the network nodes to build two SFPs. The first SFP (dashed line) includes a Firewall, an Intrusion/Malware Detection and a Video Optimizer instantiated on the two first nodes. In addition to the three VNFs of the first SFP, the second commodity (solid line) needs a special hop to a third node containing a Parental Control Function. }
\label{fig:SFCUseCaseParental}
\end{figure*}

\begin{figure}
\centering
\includegraphics[width=58.99mm]{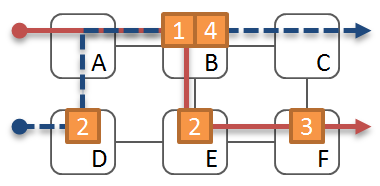}
\caption{SFCs and their SFP allocation. In this example, the solid SFP has SFC $[1;2;3]$ and the dashed SFP has SFC $[2;1;4]$. This allocation could be improved in terms of VNF costs by mutualizing VNF~2 on node $E$, and the dashed SFP would become: $\mathrm{in}\to D\to E\to B\to C \to \mathrm{out}.$ }
\label{fig:exampleSFC}
\end{figure}

In addition to the orders, the anti-affinity rules may therefore force routing at higher costs (or longer paths) in order to run network functions at reduced cost. We seek to tackle this trade off by analyzing the problem by means of a total financial cost. Our aim is to find an SFP for all commodities while minimizing the cost of function instantiation and traffic engineering together.  Figure~\ref{fig:SFCUseCaseParental} illustrates a concrete Service Function Chaining use case, and a simple example of SFCs and their instantiation on suitable SFPs is illustrated in Figure~\ref{fig:exampleSFC}.

\vspace{5mm}
\section{ILP Formulation}\label{ilpformulation}
In this section, we start by presenting an ILP Formulation that addresses exactly our problem.
\subsection{Model Parameters}
The network topology is modeled as a bidirected graph $G = (N,A)$. Each node $u$ of the graph corresponds to a physical switch equipped with a small data center that has a certain number $c_u \in \NN$ of cores, and each link $(u,v)$ corresponds to a wired connection between two switches with a bandwidth capacity $b_{uv}\in \RR_+$ and a delay (latency) $l_{uv}\in \RR_+$. We refer to $F$ as the set of available service functions. This set can be in general large but we remark that not all functions in $F$ need to be used for building SFPs. The set of commodities is denoted $C$ and each commodity $k \in C$ is characterized by a source $s_k\in N$, a destination $d_k\in N$, a bandwidth demand $b_k \in \RR_+$, a maximum tolerated latency $l_k \in \RR_+$ and a specific SFC $F_k\subset F$. \add{For simplicity, we assume that} each function occupies one core and has a maximum traffic rate it can process, $m_f\in \RR_+$. \add{It is however possible to support VNF-specific core requirements for different VNFs by modifying the node capacity constraints~\eqref{ncap} (see Section~\ref{ss:constraints}).}

The fixed relative orders between VNFs that take part in the SFC of a commodity $k$ are given by a ternary operator $p_k$ that maps the tuple of VNFs $(f,g)$ to the value $-1$ (resp. $1$) if $f$ appears after (resp. before) $g$ along the SFC. When no order is required between $f$ and $g$, $p_k(f,g) = p_k(g,f) =0$.  This parameter depicts a well-defined total or partial order.
The cost of routing per unit of traffic per link $(u,v)$ is given by the number $\psi_{uv}$ and the one for running VNF $f$ is $\psi_f$. 

Finally, we model anti-affinity rules as a set $S \subset F^2$ which contains the (unordered) tuple $(f,g)$ if and only if these two VNFs cannot handle the same service chain on the same node: hence, if they coexist on the same node, then there are at least two different commodities that are assigned to exclusively one of them on that node.

\add{The set $F$ may contain several configurations of the same VNF type if one wants to model different setup states which make some SFPs unable to share the same VNF instance. For instance, if SFP $k$ and SFP $k'$ need to meet a certain VNF $f$, with a setup $s$ and $s'$ respectively, that cannot be provided by the same instantiation, then each of these SFPs need to meet a different instantiation of the VNF $f$, each of them being in the correct setup state. Our model enforces this feature by considering the two setups as different VNF types. In the same way, this trick may permit the service provider to feature commodity isolation by exclusively allocating a commodity to a VNF instance.}
 
For more clarity, Table~\ref{tab:parameters} gives an overview of the parameters \add{and variables}.

\begin{table}
\centering
\caption{Variables and parameters}
\label{tab:parameters}
\begin{tabular}{r l l}

\hline
Parameters & & \\
\hline
$s_k$ & source node of commodity $k$ & \\
$d_k$ & destination node of commodity $k$ & \\
$b_k$ & demand of commodity $k$ & \\
 $c_u$& number of cores of node $u$ & \\
 $m_f$& maximum rate for function $f$ & \\
 $p_k(f,g)$& 1 (resp. $-1$) if $f$ appears before (resp. after) $g$ & \\
 &  on SFC of commodity $k$, 0 otherwise & \\
 $b_{uv} $& bandwidth of link $(u,v)$& \\
 $l_{uv}$& latency of link $(u,v)$& \\
 $l_k$& maximum tolerated latency for commodity $k$ & \\
$\psi_{uv}$ & cost per unit of traffic on link $(u,v)$ & \\
 $\psi_f$& cost of running function $f$ & \\
\hline
Variables & & Within \\
 \hline
$Y_u^{kf}$ & $1$ if $k$ meets its assigned $f$ before or on node $u$& Binary\\
$y_u^{kf} $ & $1$ if function $f$ handles packet $k$ on node $u$ & Binary \\
$x^f_u $& number of $f$  located on node $u$& Integer \\
$q_{uv}^k$& $1$ if path for $k$ contains link $(u,v)$ &Binary \\

\hline
\end{tabular}
\end{table}

\subsection{Decision Variables}

We introduce the ILP exact formulation of the VNF-PR problem with partial order and anti-affinity rules. We denote by $q^k_{uv}$ the routing binary variables that state whether link $(u,v)$ is used for the path of commodity $k$ or not. The number of required instantiations of VNF $f$ on node $u$ is denoted by $x^f_u \in \NN$, and we denote by $y_u^{kf}$ the binary variable stating whether VNF $f$ is assigned to commodity $k$ on node $u$ or not. To address the order constraints, we define an auxiliary variable $Y_u^{kf}$ that accounts for the state of the commodity:   $Y_u^{kf} = 1$ when on node $u$, the commodity has already been assigned to $f$ at some point along its route to $u$ {or on $u$ itself}. 

We will represent the orders between VNFs with orders between values of this variable.

\subsection{Constraints}
\label{ss:constraints}
We now turn to the formulation of our ILP's constraints. The first constraints are the classic Capacitated Multi-Commodity Flow constraints, enforcing conservation laws of the flows (\ref{mcf}), the non-violation of the bandwidth limitations of each link (\ref{lcap}), and the delay maximum tolerance (\ref{latency}):

\begin{equation}
\label{mcf}
\setlength{\nulldelimiterspace}{0pt}
\begin{split}
\sum_{(u,v)\in A} q_{uv}^k - \sum_{(v,u)\in A} q_{vu}^k =\left\{ \begin{IEEEeqnarraybox}[\relax][c]{l' s}  1 & if  $u = s_k$  \\ -1 & if  $u = d_k$ \\ 0 & otherwise   \end{IEEEeqnarraybox} \right. \quad
 \forall k\in C\quad \forall u\in N   
\end{split}
\end{equation}
\begin{equation}
\label{lcap} 
\sum_{k\in C} q_{uv}^k b_k \leq b_{uv}\quad  \forall (u,v) \in A    
\end{equation}
\begin{equation}
\label{latency}
\sum_{(u,v)\in A} q_{uv}^k l_{uv}  \leq l_k  \quad  \forall k\in C.   
\end{equation}

The following constraints state that a node cannot support more VNFs than its number of available cores (\ref{ncap}), and that each function instantiation has a limited capacity and needs to be duplicated enough to handle all commodities that are assigned to it on the corresponding node (\ref{funcapacity}):
\begin{equation}
\label{ncap}
\sum_{f\in F} x_u^f  \leq c_u\quad  \forall u\in N    
\end{equation}
\begin{equation}
\label{funcapacity}
\sum_{k\in C} y_u^{kf} b_k   \leq m_f x_u^f  \quad \forall u\in N\quad\forall f\in F.   
 \end{equation}

The anti-affinity rules modeled by the set $S$ can be formulated as follows: given a tuple of two functions that belongs to $S$, no commodity can be assigned to both of them on the same location.

\begin{equation}
 y_u^{kf} + y_u^{kg} \leq 1  \quad\forall k\in C\quad\forall (f,g)\in S\cap F_k^2 \quad\forall u\in N.    
\end{equation}

The following constraints allow one to properly impose the orders within each SFC. Moreover, one can easily establish that they also avoid the creation of cycles along a path:

\begin{equation}
\label{Ysource}
 Y_{s_k}^{kf} = 0   \quad\forall k \in C\quad\forall f \in F 
\end{equation}
\begin{equation}
\label{Ydest}
 Y_{d_k}^{kf} = 1\quad\forall k\in C \quad \forall f\in F_k
\end{equation}
\begin{equation}
\label{functionactivation}
\begin{split}
 (q_{uv}^k -1) + (Y_v^{kf} - Y_u^{kf}) \leq y_v^{kf}\quad
 \forall k\in C \quad \forall f\in F_k\quad  \forall (u,v)\in A
\end{split}
\end{equation}
\begin{equation}
\begin{split}
\label{orders}
 (Y_u^{kf} - Y_u^{kg})  p_k(f,g) \geq 0 \quad
\forall k \in C \quad\forall u\in N \quad \forall f,g \in F_k.
\end{split}
\end{equation}

Constraints (\ref{Ysource}) and (\ref{Ydest}) state that at the source, a commodity has met none of the functions, whereas at the destination, it has met all the functions included in its SFC, respectively. Constraint (\ref{orders}) accounts for the order of the functions within the same SFC: when a function $g$ appears before $f$ on the same SFC (therefore, $p_k(f,g) = -p_k(g,f)= -1$), at each node, we should have $Y_u^{kf} \leq Y_u^{kg}$. This constraint becomes trivial when no order is given between two functions of an SFC. According to constraint (\ref{functionactivation}), when the link $(u,v)$ is used - otherwise the constraint is always feasible - for commodity $k$, the commodity meets its assigned copy of VNF $f$ on $v$ if its state variable $Y_v^{kf}$ jumps from 0 to 1. Constraint (\ref{Ydest}) ensures that this jump happens once, and constraint (\ref{orders}) ensures that the jumps occur in the correct order given by $p_k$. Those four constraints therefore give a profile of the $Y$ variable that along each chosen path reads the state of the commodity with respect to function traversal constraints.

Moreover, in order to improve the linear relaxation (LP) bound, we add the following constraints:

\begin{equation}
\label{newdisjunction}
y_u^{kf}+y_u^{kg} \leq \sum_{(v,u) \in A} q_{vu}^k \quad \forall k\in C\quad\forall u\in N\quad \forall (f,g)\in S
\end{equation}
\begin{equation}
\label{nocheatassignment}
y_u^{kf} \leq \sum_{(v,u) \in A} q^k_{vu} \quad \forall f\in F \quad \forall u\in N\quad \forall k\in C
\end{equation}

\begin{equation}
\label{totalassignment}
\sum_{u\in N} y_u^{kf} =\left\{ \begin{IEEEeqnarraybox}[\relax][c]{l' s}  1 & if  $f \in  F_k$  \\  0 & otherwise   \end{IEEEeqnarraybox} \right.\quad \forall k \in C\quad \forall f \in F
\end{equation}
\begin{equation}
\label{cflclassic}
y_u^{kf}\leq x_u^f \quad\forall u\in N\quad\forall f\in F\quad\forall k\in C.
\end{equation}

Constraint~(\ref{newdisjunction}) specifies the anti-affinity rules in terms of fractional rates: no traffic is assigned in its totality to two incompatible VNFs on the same location. In turn, constraint~(\ref{nocheatassignment}) states that an assignment can be made on a node only if the commodity passes through this node. Constraint~(\ref{totalassignment}) sets the total value of a required assignment to $1$ to avoid unnecessary additional assignments. Finally, constraint~(\ref{cflclassic}) ensures that no assignment can be made if a VNF is not instantiated. It is redundant with constraint~(\ref{funcapacity}) but this kind of cut is well known to improve the LP-bound.

\subsection{Formulation}
Our aim is to minimize the total financial cost encompassing the routing ($\psi_{uv}$) and VNF ($\psi_f$) costs. Therefore, we formulate the VNF Placement for Service Chaining with Partial Orders and Anti-Affinity Rules ILP as the following:
$$\displaystyle \min \sum_{(u,v)\in A, k\in C}q^k_{uv} b_k \psi_{uv} + \sum_{u\in N, f\in F} x_u^f \psi_f  $$
$$\mathrm{s.t. }\quad  (1) - (14)$$
$$ q^k_{uv} , y^{kf}_u, Y_u^{kf}\in \{0,1\}, x_u^f \in \NN.$$

\subsection{Handling Commodity Rejection}

Our model formulation is flexible enough to easily incorporate rejection decisions. First of all, this feature allows one to always have a non-empty admissible set that consists in artificial solutions where commodities are allocated and others are rejected. This variation can also respond to situations where resources are congested and a decision has to be made on which commodities to admit in the network and which ones to reject. In this work, we use this feature to handle complicated SFCs (see \ref{subsec:algorithm}) for which the fractional optimal solution does not provide a feasible SFP.  To provide this option, we introduce a binary decision variable, $\{\bar{q}_k\}_{k\in C}$, that equals $1$ when a commodity is rejected or $0$ when it is accepted in the network. This variable will be added to the flow conservation constraints to allow a certain commodity $k$ flow to be identically zero: 

\begin{equation}
\setlength{\nulldelimiterspace}{0pt}
\begin{split}
\sum_{(u,v)\in A} q_{uv}^k - \sum_{(v,u)\in A} q_{vu}^k =\left\{ \begin{IEEEeqnarraybox}[\relax][c]{l' s}  1 - \bar{q}_k & if  $u = s_k$  \\  \bar{q}_k - 1 & if  $u = d_k$ \\ 0 & otherwise   \end{IEEEeqnarraybox} \right. \\
\forall k\in C\quad \forall u\in N.
\end{split}  
\end{equation}

In addition, when a commodity is rejected, there is no VNF assigned to it:

\begin{equation}
y_u^{kf}, Y_u^{kf} \leq 1-\bar{q}_k\quad \forall k\in C\quad \forall f\in F \quad\forall u\in N
\end{equation}

and constraint~(\ref{Ydest}) is replaced with constraint~(\ref{totalvalueY}):
\begin{equation}
\label{totalvalueY}
Y_{d_k}^{kf} = 1-\bar{q}_k \quad \forall f\in F_k\quad \forall k\in C.
\end{equation}

In turn, we replace the constraint~(\ref{totalassignment}) by the following:
\begin{equation}
\label{totalvaluey}
\sum_{u\in N} y_u^{kf}= \left\{ \begin{IEEEeqnarraybox}[\relax][c]{l' s}  1-\bar{q}_k & if  $f \in  F_k$  \\  0 & otherwise   \end{IEEEeqnarraybox} \right. \quad \forall f\in F\quad \forall k\in C.
\end{equation}

The new ILP thus reads:

\begin{equation}
\label{ILP}
 \min \sum_{(u,v)\in A, k\in C}q^k_{uv} b_k \psi_{uv} + \sum_{u\in N, f\in F} x_u^f \psi_f +r\sum_{k\in C} \bar{q}_k  b_k 
\end{equation} 
$$\mathrm{s.t.}\quad  (2) - (7), (9) - (12), (14) - (18)$$
$$\bar{q}_k, q^k_{uv},y^{kf}_u, Y_u^{kf}\in \{0,1\}, x_u^f \in \NN,$$

where $r$ is a sufficiently large parameter, for instance: $$  r > \frac{ \max b_k |A| |C| \psi_{uv} +|F| |N|\max c_u \max \psi_f}{\min b_k} .$$

From now on, we always consider this variant of the problem.

\vspace{5mm}
\section{Greedy algorithm}\label{greedyalgorithm}
We first illustrate a simple greedy algorithm (Algorithm~\ref{alg:greedy}) that we will compare with the LP-based heuristic approach proposed in the next section. The greedy approach treats commodities in a sequential fashion: it allocates the commodities on the network one-by-one, while instantiating VNFs on-the-fly that can remain available for subsequent demands. Of course, the algorithm may reject commodities when their allocation is infeasible. It is simple to formulate an ILP problem that serves our needs and that treats commodities in sequence: we define an additional parameter that accounts for the actual use rate of VNFs on each location given previous allocations. At each step, the algorithm can decide to either instantiate new VNFs or use already existing ones, depending on which is less costly. Let $\{r_u^f\}$ be this family of parameters. Constraint (5) then becomes for step $k$, for each node $u$ and each VNF $f$: 

\begin{equation}
 y_u^{kf} b_k   +r_u^f \leq m_f x_u^f.
\end{equation}

This rate has to be updated at each step output.

\begin{algorithm}
\caption{Greedy}
\label{alg:greedy}
\begin{algorithmic}[1]
\Require Network Topology $(N,A)$, set of commodities $C$ and their SFCs
\Ensure  Integer solution $\{q_{uv}^k\}_{k,(u,v)}, \{y_u^{kf}\}_{u,k,f}, \{x_u^f\}_{u,f}$ and a rejected set $S$
\State $r_u^f = 0 \quad \forall u, \forall f$, $S = \emptyset$

\For{$k\in C$}
	\If{$k$ is feasible}
		\State allocate $k$ with ILP
	\Else
		\State $S\gets S\cup\{k\}$
	\EndIf
\State update $x$
\State update all link capacities
\State update use rates $r_u^f$ for each newly installed $f$ on each node $u$ 
\EndFor

\end{algorithmic}
\end{algorithm}

\vspace{5mm}
\section{LP-based Heuristic}\label{heuristicdesign}

To solve efficiently our problem, we propose a heuristic approach based on the LP-relaxation of the original ILP in \ref{ilpformulation}. Before delving into the details of our heuristic algorithm, we first describe in the next paragraph its main sub-routine.

\subsection{Restricted Problem: VNF Deployment on Selected Routes}

We now describe a sub-routine used by our heuristic approach, which optimizes the VNF deployment given pre-computed routes. As a by-product, the following sub-routine can be thought of as a stand-alone routine, which is useful for re-allocating all VNFs on a set of established routes, e.g., after costly deviations caused by greedy allocations of SFPs, or after a resource breakdown which would demand a new deployment. To this aim, the rejection variables are here added to the constraint~(\ref{functionactivation}) so that they always remain feasible for rejected commodities:

\begin{equation}
 (q_{uv}^k -1) + (Y_v^{kf} - Y_u^{kf}) \leq y_v^{kf} + \bar{q}_k\quad 
\end{equation}
The ILP we provide to address this situation hence reads:
 $$ \min \sum_{u\in N, f\in F} x_u^f \psi_f +r\sum_{k\in C} \bar{q}_k  b_k$$
$$\mathrm{s.t.}\quad  (4) - (7), (9) - (12), (14), (16) - (18)$$
$$  y^{kf}_u, Y_u^{kf}, \bar{q}_k \in \{0,1\}, x_u^f \in \NN.$$

\subsection{Algorithm}\label{subsec:algorithm}
We are finally ready to describe our heuristic approach to jointly compute the routing paths and the virtual functions placement with partial order and anti-affinity constraints. Algorithm~\ref{algo:heur} outlines the basic steps of our approach.

\begin{figure}

\centering
\includegraphics[width=48.99mm]{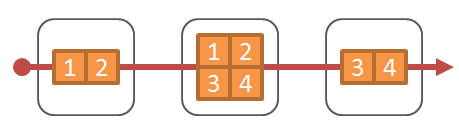}

\caption{An example of an admissible LP-solution that does not provide integer solution. In this example, the commodity's SFC is [1;2;3;4], and all couples of functions are disjoint. The LP-SFP where all assignment values equal $0.5$ is admissible but the shortest integer SFP has length $4$.} 
\label{lppath}

\includegraphics[width=58.99mm]{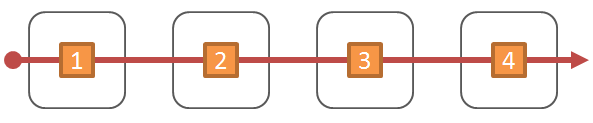}
\caption{Any feasible path allocation for the SFC in Fig.~~\ref{lppath} requires at least one hop between each VNF traversal.}
\label{ilppath}
\end{figure}

Given the optimal solution of the relaxed version of the integer problem (\ref{ILP}), we use a rounding scheme to retrieve a best candidate path for each commodity. We then deploy the VNFs optimally with respect to this path selection using the restricted problem described above. We claim that the fractional paths given by the LP relaxation are good candidates for feasible SFPs. However, this may not necessarily be the case for all SFCs. A complicating combination of anti-affinity rules and precedence constraints may blur the LP evaluation of the feasible minimal number of hops for a commodity's unsplittable SFP. We give a simple example of a complicating order and anti-affinity combination on an SFC in Figures~\ref{lppath} and \ref{ilppath} to illustrate this situation. In those cases, we retrieve rejected commodities and allocate them at the end of the procedure with the greedy algorithm. The concerned commodities can then be allocated on a path built on new VNF instantiations, already existing VNFs, or both. 

As a result, the validity of the LP solution is a crucial question that deserves commenting on. The linearized variables $q, \bar{q}, y$ and $Y$ keep their definitions, except they always account for the fraction of the traffic given by their value. In particular, for $k\in C$, $f\in F_k$, $u\in N$, $y_u^{kf}$ defines the portion of the commodity $k$ that is assigned to VNF $f$ on node $u$, and $Y_u^{kf}$ defines the fraction of commodity $k$ that has already met VNF $f$ during its route to $u$. As such, they do not entirely enforce anti-affinity rules and orders. One could further add linear cuts in order to improve the quality of the relaxation, but we remark that this has the tendency to considerably increase the solver execution time for large scale instances. Moreover, we argue that complicated overlapping of anti-affinity rules and precedence constraints should be relatively rare and can be considered as particular cases for which it is acceptable to handle a first rejection at the end of the procedure. We remark that SelectPath procedure, at line \ref{algo:heur:path} of Algorithm \ref{algo:heur}, stands for any rounding procedure that converts a fractional solution into a feasible path over the capacitated graph. In our numerical evaluations, for instance, we implemented SelectPath as a deterministic walk that, for each commodity, starts at the source and adds the (not yet visited) link with highest traffic rate to the path until it reaches the destination. For example, in \cite{even2016approximation}, the authors form an unsplittable SFP from the fractional optimum by following a random walk upon the support of the fractional SFP.

\begin{algorithm}[t]
\caption{Heuristic}
\label{algo:heur}
\begin{algorithmic}[1]
\Require  Network Topology $(N,A)$, set of commodities $C$ and their SFCs 
\Ensure  Integer solution $\{q_{uv}^k\}_{k,(u,v)}, \{y_u^{kf}\}_{u,k,f}, \{x_u^f\}_{u,f}$ and a rejected set $S$

\State $S = \emptyset$
\State $\hat{q}^k_{uv} \gets$ route values from LP-optimal solution (ILP~(\ref{ILP}))
\State $q\gets$ SelectPath($\hat{q}$) \label{algo:heur:path}
\State Solve ILP for VNF deployment on selected routes with $q$, $C$
\For{$k\in C$}
	\If{$k$ is allocated}
	\State update  $y_u^{kf}$, $f\in F$, $u\in N$
	\Else \textit{ ($k$ is rejected)}
	\State set $q_{uv}^k = 0 \quad\forall (u,v)\in A$
	\State $S\gets S\cup\{k\}$
	\EndIf
\EndFor
\State define use rates as $r_u^f =  \sum_{ \mathrm{k\notin S}} y_u^{kf} b_k$, $f\in F$, $u\in N$
\State update all network capacities
\State run Greedy on $S$ with initialized use rates $r_u^f$
\State update $q$, $y$, $x$ and $S$

\end{algorithmic}
\end{algorithm}

\section{Performance Evaluation}\label{evaluation}

This section presents the simulation setup, experimental results and overall observations.

\subsection{Simulation Setup}\label{ss:setup}
\begin{figure*}
\centering
\includegraphics[width=\textwidth]{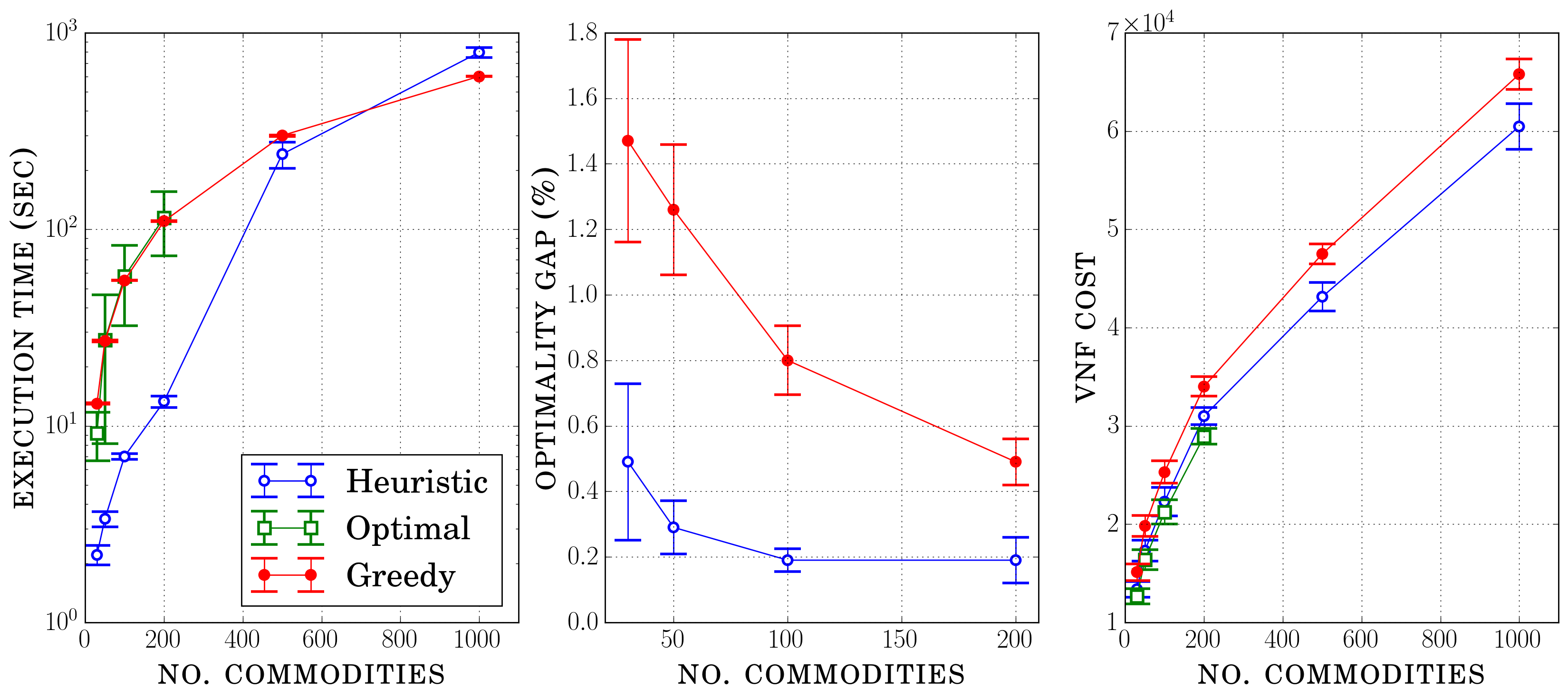}
\caption{Scalability in terms of number of commodities.}
\label{commodities}
\end{figure*}

We ran all our tests using as solver IBM ILOG CPLEX 12.6 and developed the non optimization parts of our algorithms in Python. Each point we represented corresponds to the average value for \add{15 instances} generated on the GEANT topology which contains 22 nodes and 36 bidirected edges. For all our instances, the nodes each had a total of 20 available cores and the bandwidth limit of the links were each 40Gb, while the VNFs capacities were fixed to 5Gb. The average traffic cost per unit per link was fixed to 10\$/Mb/month and the average price of running a VNF was either 200\$/month or 1000\$/month. The ratio between the two costs, denoted $s$, is in the former case 5\%, and in the latter case 1\% (costly VNFs). Those prices were generated uniformly within an interval of $\pm$ 20\%.  \add{In all our instances, the commodity sources and destinations were chosen uniformly at random amongst the network nodes.} In a first experiment, we show the scalability of the LP-based heuristic approach in terms of the number of commodities to allocate. Next,  we point out its efficiency as we stress the orders and anti-affinity rules. All instances feature SFCs of lengths varying from 4 to 8 VNFs generated uniformly from a total of 10 existing VNFs. We only consider small demands from 100 to 500Mb so that no network resource is stressed in any of the instances. Indeed, in this work, we focus on the sole effects of orders and anti-affinity rules as we stress them.

\subsection{Results}

\begin{figure*}[t]
\centering
\includegraphics[width=\textwidth]{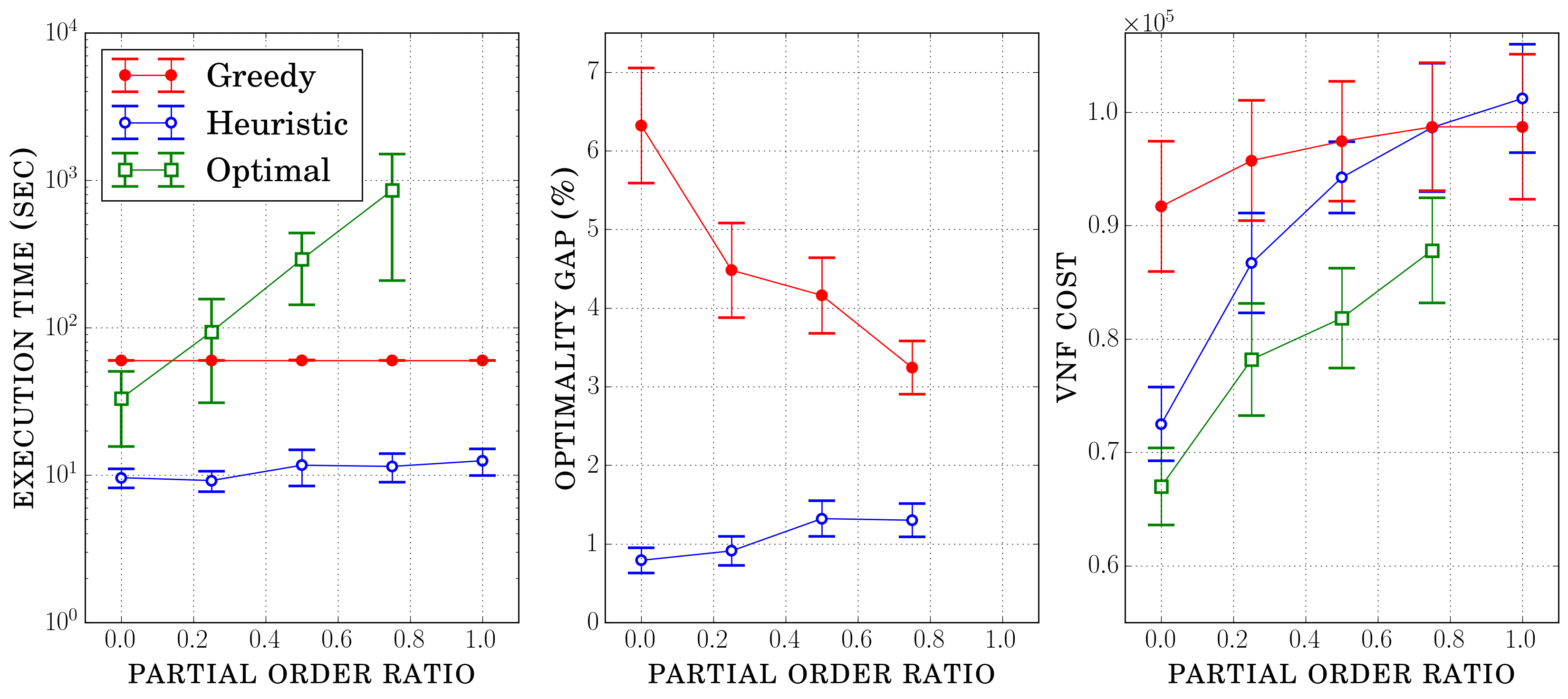}
\caption{Order constraints influence.}
\label{partial_order}
\end{figure*}

\begin{figure*}[tb]
	\centering
	\includegraphics[width=\textwidth]{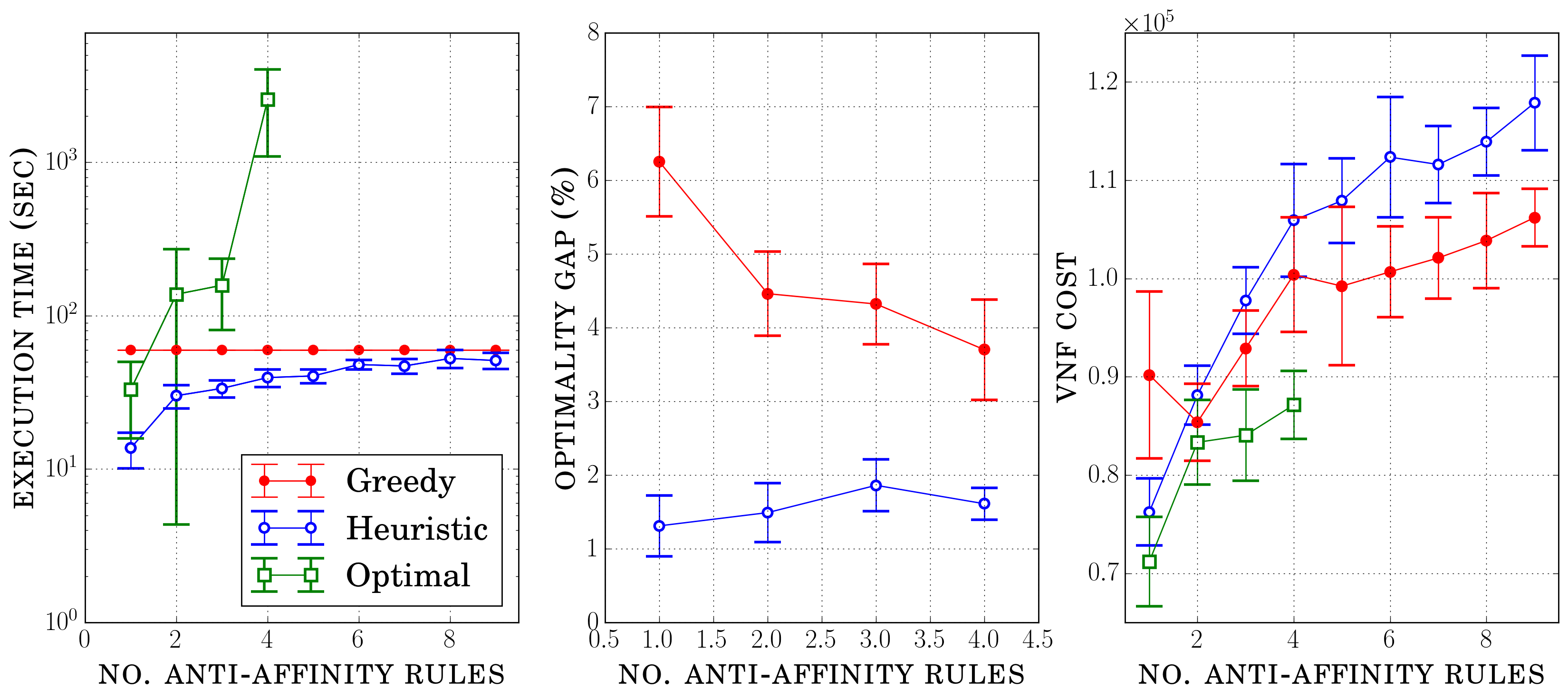}
	\caption{Anti-affinity rules influence.  }
	\label{anti_affinity}
\end{figure*}

We now comment upon the results of our numerical investigations, shown in Figures~\ref{commodities} -- \ref{anti_affinity}. Each figure compares the execution time, the optimality gap and the total cost of VNFs in different scenarios \add{and present error bars accounting for an estimated 95\%-confidence interval calculated following the $t$-distribution model}.

In Figure~\ref{commodities}, the instances feature partially ordered SFCs with 6 anti-affinity rules generated randomly. The cost ratio $s$ (see \ref{ss:setup}) is fixed to 5\%. According to Figure~\ref{commodities}, the LP-based heuristic can solve the instances with up to $1000$ commodities whereas the ILP cannot manage all instances with $500$ or $1000$ commodities within reasonable time. Our time performances remain acceptable up to $200$ commodities whereas the optimality gaps do not exceed $2\%$. 

In the next experiments (Figures~\ref{partial_order} and \ref{anti_affinity}), more importance is given to the VNF costs by setting $s$ to 1\% and we remark that this increases considerably all solver times. \add{In order to keep the optimal value tractable, at least for small parameters, we fix the number of commodities to 100.}

In Figure~\ref{partial_order}, we first analyze the performances as the number of order constraints increases. We use the ``partial order'' parameter $\theta = \frac{2|\mathrm{supp}(p_k)|}{|F_k|(|F_k| -1)}$\footnote{Here $\mathrm{supp}(p)$ denotes the support of the function $p$, that is, the set over which $p$ is non-zero.}, in other words the number of precedence constraints over the total number of VNF couples within each SFC and plot the results for $\theta = 0 \mbox{ (no order)}, 0.25, 0.5, 0.75, 1$ (total order). The LP-based heuristic solves all instances within seconds whereas the ILP execution time quickly explodes. \add{Regarding the VNF costs, one can remark that while the two algorithms, Greedy and the LP-based heuristic, seem to instantiate more VNFs, the optimality gaps remain quite satisfactory, especially for the LP-based heuristic. Indeed, the VNF cost is only a part of the total cost to minimize, the other part being the routing price, which the LP-based heuristic seems to optimize better.} 
 
Finally, we vary the number of anti-affinity rules and analyze the performance of the LP-based heuristic on the exact same scenario. Despite the low ratio $s = 1\%$, we observe in Figure~\ref{anti_affinity} that the LP-based heuristic instantiates more VNFs than Greedy but routes at lower costs and hence performs better. \add{The time performance indicates that the LP-based heuristic is a good solution to smooth out the sensitivity of the problem with respect to the number of anti-affinity rules. The same remark can be made concerning the partial order parameter. }

In all scenarios, the greedy and LP-based heuristic algorithms both give acceptable objectives, but the latter always gives closer to optimal results - not exceeding the optimal value by more than 2\%. Moreover, Greedy cannot compete with the LP-based heuristic in terms of time performance. In the first experiment, Greedy is at first as slow as the ILP, as result of the numerous summoning of the solver to allocate each demand, whereas the LP-based heuristic can perform better and 4 to 8 times faster than the solver.
The solver time explodes quickly as the precedence constraints become more numerous, whereas the LP-based heuristic is from 3 to 70 times faster and steady as the number of those constraints increase. The same phenomenon is observable when we increase the number of anti-affinity rules, except that the growth of execution time for the LP-based heuristic corresponds to 1) the growth of LP time as the number of anti-affinity rules grows, 2) the more numerous commodities that are allocated greedily at the end of the procedure. \add{Because of the solver's inability to solve some instances when they become large or highly constrained, the optimality gaps  in all experiments are not always available.}

\subsection{Overall Observations}

In a high capacitated network in terms of link and nodes resources, the LP-based heuristic performs very close to optimum with a competitive time that beats the greedy approach in all the instances that we created up to 1000 commodities.  Through all runs, the crushing majority of the LP-based heuristic's execution time is used by the LP step. Onto the same scenario, we observe three stressing factors along which the execution time explodes: the number of precedence constraints within SFCs, the number of anti-affinity rules, and the cost ratio $s$. Beyond time performances, the numerical results show, as intuitively expected, that more flexibility in the orders can remarkably reduce the VNF costs, while anti-affinity rules force us to instantiate more VNFs. Our heuristic nevertheless smooths out the two difficulties and solves - with reduced and steady time - instances that the ILP cannot solve within reasonable time. Of course, those factors are not the only ones. The case where the node resources are stressed is still problematic.  Nonetheless, the LP relaxation should be accurate enough to handle commodity rejection on networks with stressed link resources.

\section{Conclusion and Perspectives}\label{conclusion}

In this paper we tackled the joint optimization problem consisting in placing Service Functions (SF) and routing multiple commodities over a general capacitated network. Each commodity requires handling by a set of SFs with \emph{partial} order constraints. The Virtual Machines (VMs) running virtual functions have limited processing capabilities, while the nodes hosting VMs are characterized by limited memory. Moreover, \emph{anti-affinity} rules forbid the assignment of certain VNFs to one commodity on the same location, in order to ensure that SF paths (SFPs) do not entirely break down by a single server failure.

Since the problem is NP-hard, we proposed a heuristic approach based on the relaxation of the optimal integer linear programming (ILP) formulation. We added redundant linear constraints to the original ILP formulation that permit us to reduce 
the optimality gap.

We evaluated the performance of our heuristic approach by benchmarking it against both the optimal ILP solution and a greedy method that processes the commodities one by one. We demonstrated numerically that our heuristic algorithm strikes a good balance among optimality gap, VNF cost and execution time.

In terms of future development, we plan to refine our heuristic approach by boosting its performance in the presence of stressed node resources. Moreover, we envision adapting our heuristic to be able to better accommodate in an online fashion the arrival of new commodities, while still outperforming the greedy approach.

\bibliographystyle{apa}

\end{document}